\documentclass[sigconf, screen]{acmart}
\usepackage{colortbl}
\usepackage{multirow}
\usepackage{soul}
\PassOptionsToPackage{dvipsnames}{xcolor}

\hypersetup{
     colorlinks=True,
     linkcolor=blue,
     filecolor=blue,
     citecolor = blue,      
     urlcolor=cyan,
 }

\newcolumntype{P}[1]{>{\centering\arraybackslash}p{#1}}
\definecolor{ForestGreen}{RGB}{34,139,34}
\definecolor{RubineRed}{RGB}{209,0,86}
\definecolor{White}{HTML}{FFFFFF}
\definecolor{Black}{RGB}{0,0,0}

\newcommand{\cmark}{\color{ForestGreen}{\checkmark}}
\newcommand{\xmark}{\color{RubineRed}{$\times$}}

\newcommand{\headerline}{\rowcolor[HTML]{4472C4}}
\newcommand{\greenrow}{\rowcolor[HTML]{71da71}}
\newcommand{\yellowrow}{\rowcolor[HTML]{ffff70}}
\newcommand{\lightrow}{\rowcolor[HTML]{CFD5EA}}
\newcommand{\darkrow}{\rowcolor[HTML]{E9EBF5}}

\newcommand{\header}[1]{\cellcolor[HTML]{4472C4}{\color{White}{\textbf{#1}}}}
\newcommand{\headerb}[1]{\color{White}{\textbf{#1}}}

\newcommand{\darkcell}[1]{\cellcolor[HTML]{CFD5EA}{\color{Black} #1}}
\newcommand{\ncell}[1]{#1}

\newcommand\logicalor[1]{{\fontfamily{cmss}\selectfont #1}}
\newcommand{\takeaway}[1]{
    \noindent\fbox{%
        \parbox{.96\columnwidth}{%
            \textbf{Takeaway}: {#1}
        }%
    }
}

\AtBeginDocument{%
  \providecommand\BibTeX{{%
    \normalfont B\kern-0.5em{\scshape i\kern-0.25em b}\kern-0.8em\TeX}}}

\setcopyright{acmcopyright}
\copyrightyear{2021} 
\acmYear{2021} 
\acmConference[ARES 2021]{The 16th International Conference on Availability, Reliability and Security}{August 17--20, 2021}{Vienna, Austria}
\acmBooktitle{The 16th International Conference on Availability, Reliability and Security (ARES 2021), August 17--20, 2021, Vienna, Austria}
\acmPrice{15.00}
\acmDOI{10.1145/3465481.3470065}
\acmISBN{978-1-4503-9051-4/21/08}

\begin{document}

\title{On the Evaluation of Sequential Machine Learning for Network Intrusion Detection}

\author{Andrea Corsini}
\email{andrea.corsini@unimore.it}
\affiliation{%
  \institution{University of Modena and Reggio Emilia}
  \streetaddress{Via Pietro Vivarelli}
  \city{Modena}
  \country{Italy}
  \postcode{41125}
}
\author{Shanchieh Jay Yang}
\email{jay.yang@rit.edu}
\affiliation{%
  \institution{Rochester Institute of Technology}
  \streetaddress{8 Lomb Memorial Drive}
  \city{Rochester, NY}
  \country{USA}
  \postcode{14623}
}

\author{Giovanni Apruzzese}
\email{giovanni.apruzzese@uni.li}
\affiliation{%
  \institution{University of Liechtenstein}
  \streetaddress{Fürst-Franz-Josef-Strasse}
  \city{Vaduz}
  \country{Liechtenstein}
  \postcode{9490}
}

\begin{abstract}
Recent advances in deep learning renewed the research interests in machine learning for Network Intrusion Detection Systems (NIDS).
Specifically, attention has been given to sequential learning models, due to their ability to extract the temporal characteristics of Network traffic Flows (NetFlows), and use them for NIDS tasks.
However, the applications of these sequential models often consist of transferring and adapting methodologies directly from other fields, without an in-depth investigation on how to leverage the specific circumstances of cybersecurity scenarios; moreover, there is a lack of comprehensive studies on sequential models that rely on NetFlow data, which presents significant advantages over traditional full packet captures.
We tackle this problem in this paper. We propose a detailed methodology to extract temporal sequences of NetFlows that denote patterns of malicious activities. Then, we apply this methodology to compare the efficacy of sequential learning models against traditional static learning models. In particular, we perform a fair comparison of a `sequential' Long Short-Term Memory (LSTM) against a `static' Feedforward Neural Networks (FNN) in distinct environments represented by two well-known datasets for NIDS: the CICIDS2017 and the CTU13. Our results highlight that LSTM achieves comparable performance to FNN in the CICIDS2017 with over 99.5\% F1-score; while obtaining superior performance in the CTU13, with 95.7\% F1-score against 91.5\%. This paper thus paves the way to future applications of sequential learning models for NIDS.
\end{abstract}



\maketitle

\section{Introduction}
\label{sec:introduction}

Network Intrusion Detection Systems (NIDS) are of paramount importance for the protection of network infrastructures. Many detailed works (e.g.,~\cite{mishra2018detailed, survey2013, survey2016, survey2019}) highlighted the significant advances of NIDS, which are subject of continuous improvements. Indeed, NIDS must face significant challenges such as the ever-changing network environments as well as skilled and motivated adversaries. In particular, the recent promising successes in machine and deep learning renewed the interest in devising autonomous defensive systems~\cite{survey2016, survey2019}, giving life to a next generation of NIDS.

Among the plethora of proposals that exploit Machine Learning (ML) techniques, one promising direction seeks to leverage \textit{sequential learning} methods that rely on the enticing idea that specific network events exhibit temporal patterns. When properly applied, sequential ML methods---and especially their ``deep'' variants---offer the appreciable opportunity to automatically extract these temporal patterns. These patterns can then be used to create conventional anomaly- or signature-based detection systems. 
Despite abundant literature, an effective and realistic deployment of sequential learning models for NIDS is yet a distant goal. For instance, identifying temporal patterns is challenging because many factors must be considered, such as, the adopted network protocols; the web-services employed by the network; and the possible threats that may target the network. 

Moreover, the massive size and variability of modern network traffic data makes timely analyses based on traditional packet captures (PCAP) a difficult objective -- which is further aggravated by the increasing usage of encrypted traffic. To mitigate this issue, many proposals (e.g.,~\cite{flowIDS}) favor the inspection of Network Flows (NetFlows) instead of PCAP data. In this context, we make an intriguing observation: although it has been empirically shown that cyber-attacks exhibit temporal patterns that can be easily extracted at the packet level, it is less clear whether this is still true at the `higher' NetFlow level\footnote{NetFlows are a statistical summary of packets, grouped according to a set of rules~\cite{NetFlow}.}. Indeed, temporal patterns within packets are not guaranteed to exist also in the corresponding NetFlows, which are metadata that summarize the corresponding PCAP data.

In this paper we devise a novel methodology for (i) investigating the existence of temporal patterns in malicious NetFlows; and, if such patterns exist, (ii) assess their effectiveness for detection purposes.
The primary focus of past literature on sequential ML-NIDS is the proposal of solutions that ``outperform'' the state-of-the-art according to some performance metrics; however, the investigation of the actual benefits brought by these solutions is often neglected. In contrast, our objective is a preliminary analysis focused on the investigation of the existence, advantages and shortcoming brought by using temporal patterns at the NetFlow level for NIDS. To the best of our knowledge, this paper is the first effort that analyzes and performs a fair comparison of sequential ML models against their `static' counterparts.

We experimentally evaluate the proposed methodology on two well-known and recent datasets for NIDS, the CTU13~\cite{CTU13} and the CICIDS2017~\cite{CICIDS2017}. We apply our method to extract temporal patterns, and use such patterns to train and compare two Deep Learning (DL) algorithms: one based on Long-Short-Term-Memory (LSTM), which leverages sequential learning; and one based on Feedforward Neural Networks (FNN), which is a traditional `static' learning method. Our results highlight that the patterns extracted by our method yield proficient detectors in the CICIDS2017 scenario, with both LSTM and FNN achieving over 99.5\% F1-score; whereas the LSTM significantly outperforms the FNN detector in the CTU13 scenario, with 95.7\% F1-score against 91.5\%. 

We are confident that this paper will shed some light on the application of sequential learning models for NetFlow-based NIDS. Our work thus paves the way to more reliable defensive platforms, which can jointly combine existing detection techniques with novel solutions that also consider the temporal axis.

To summarize the main contributions of this work:
\begin{itemize}
    \item we devise an original method to extract temporal patterns from NetFlows that can be leveraged by sequential ML;
    \item we evaluate the proposed method to explore its benefits by performing a fair comparison of a `static' FNN against a `sequential' LSTM on two well-known datasets for NIDS.
\end{itemize}

The remainder of this paper is structured as follows. Section~\ref{sec:related} discusses related work. Section~\ref{sec:methodology} describes the proposed methodology for extracting temporal patterns in malicious NetFlows. Section~\ref{sec:experiments} presents the experimental settings. Section~\ref{sec:results} evaluates our proposed method. Section~\ref{sec:conclusions} concludes the paper with final remarks and future work.

\section{Related Work}
\label{sec:related}

There exist many proposals for detecting malicious traffic by means of machine learning techniques (e.g.,~\cite{survey2016, mishra2018detailed}). There is also abundant literature on time-based analyses for NIDS (e.g.,~\cite{MalHosts, zhou2003mining}). 

This paper lies at the intersection of these two application
fields. Our objective is investigating the effectiveness of temporal relationships for ML-based NIDS (ML-NIDS), hence our literature review focuses on those proposals that leverage machine- and, specifically, deep-learning by considering also the temporal axis, usually through Recurrent Neural Networks (RNN).

Among the earliest successful applications of sequential learning models to NIDS there is the 2017 proposal in~\cite{FirstRNN}.
Since then, several works have proposed learning models that extract temporal relationships in the network traffic and use them for ML purposes. We categorize and discuss such works on the basis of the approach used to extract the temporal patterns, namely: \textit{automatic} approaches, where the temporal patterns are extracted through a pipeline that inspects the input data and automatically learns the most significant temporal features; \textit{manual} approaches, where the temporal patterns are extracted exclusively on the basis of the input features.

We provide an overview of the major differences among related work in Table~\ref{tab:works}. This table shows the datasets considered, the employed DL techniques (CNN stands for Convolutional Neural Networks), and whether the approach is tailored for anomaly- or misuse-based\footnote{Note that within the `misuse' category we also include approaches that use the trained ML models to perform the detection.} NIDS.

\begin{table}[!htbp]
\centering
\caption{Overview of related work on sequential ML-NIDS.}
    \resizebox{\columnwidth}{!}{
    \begin{tabular}{c|c|c|c|c|c|c}
    \toprule
    & \textbf{\textit{Paper}}
    & \textbf{\textit{Datasets}}
    &\textbf{\textit{
        \begin{tabular}{c} Detection \\ method \end{tabular}}
        }
    &\textbf{\textit{CNN}}
    &\textbf{\textit{RNN}}
    &\textbf{\textit{FNN}}\\
    \midrule
    \midrule
     \parbox[t]{2mm}{\multirow{5}{*}{\rotatebox[origin=c]{90}{\textit{Automatic}}}} 
    & \cite{STIDM} & ISCX2012, CICIDS2017 & misuse & \cmark & \cmark & \xmark \\
    & \cite{DLFlowSummary} & CTU13, ISOT & misuse & \cmark & \cmark & \cmark \\
    & \cite{DeepHierarchicalNet} & CICIDS2017, CTU13 & anomaly & \cmark & \cmark & \xmark \\
    & \cite{GraphLSTM} & CTU13 & misuse & \xmark & \cmark & \xmark \\
    & \cite{LuNet} & NSLKDD, UNSW-NB15 & misuse & \cmark & \cmark & \xmark \\
    \midrule
    \midrule
    \parbox[t]{2mm}{\multirow{5}{*}{\rotatebox[origin=c]{90}{\textit{Manual}}}} & \cite{AggregationRules} 
    & CICIDS2017  & anomaly & \xmark & \cmark & \cmark \\
    & \cite{AnomalyRNN} & ISCX2012  & anomaly & \xmark & \cmark & \cmark \\
    & \cite{NIDSLSTMEmbedding} & UNSW-NB15  & misuse & \xmark & \cmark & \xmark \\
    & \cite{BotnetVAE} & CTU13 & anomaly & \xmark & \cmark & \xmark \\
    & \cite{BotnetRNN} & CTU13 &  misuse & \xmark & \cmark & \xmark \\
    \bottomrule
    \end{tabular}
    }
\label{tab:works}
\end{table}

We do not consider works whose proposal is evaluated exclusively on the (outdated and widely deprecated~\cite{CICIDS2017}) NSLKDD dataset (e.g.~\cite{boukhalfa2020lstm}), and that have been published recently (since 2017). Finally, although many works use RNN for NIDS (e.g.,~\cite{khan2019scalable}), they often neglect considering temporal dependencies, hence we do not consider such works in our analysis.

\subsection{Automatic extraction of features}
\label{sec:automatic}
These works rely on an end-to-end pipeline that leverages spatio-temporal dependencies, usually by means of CNN and RNN. The CNN automatically learns both spatial and temporal features, thus removing the need of manually extracting the features used to devise the detector. In some of these proposals (i.e.,~\cite{LuNet, DeepHierarchicalNet}), the network traffic is inspected at the packet level by transforming each packet into an image.
These images are first spatially analyzed through a CNN, and then, the feature maps describing packets are used to extract temporal patterns by means of sequential models.

The authors of~\cite{STIDM} proposed a Spatial and Temporal Aware Intrusion Detection Model (STIDM) composed of LeNet-5\footnote{LeNet-5 is a well-known CNN architecture: \url{http://yann.lecun.com/exdb/lenet}} and a modified version of the LSTM. LeNet-5 was used to extract spatial features, while the modified LSTM was used to extract temporal patterns by taking into account the time intervals between packet exchanges.
STIDM was trained on truncated flows of multiple packets created by grouping on the tuple (\textit{SourceIP, SourcePort, DestinationIP, DestinationPort, Protocol}). We observe that it is difficult to attribute the outstanding performance of STIDM to the CNN and/or the LSTM from the results shown in~\cite{STIDM}. Furthermore, STIDM works on truncated sequences of packets, which makes determining the existence of temporal patterns in NetFlows unfeasible.

The approach in~\cite{DeepHierarchicalNet} is similar to~\cite{STIDM}, as they use the same STIDM structure to extract spatial features and temporal patterns, but the LSTM architecture is different.
In this work, the overall model was used to analyze sequences of packets composed of only the first 160 bytes of every packet; then, the packets are aggregated by following the same five-tuple of STIDM, but only the first 10 packets in each sequence are considered.
Similarly to~\cite{STIDM}, from~\cite{DeepHierarchicalNet} it is difficult to conclude anything about the existence temporal patterns among NetFlows and about the true benefits brought by the LSTM, since the results of the tested models (CNN, LSTM and CNN+LSTM) showed comparable performance.

A different proposal for a end-to-end pipeline is in~\cite{LuNet}. Here, instead of chaining a CNN with a LSTM (as in~\cite{STIDM}), the authors first create a block composed by these two techniques, and then stack many of these blocks obtaining higher abstraction levels. Although the final performance of such architecture achieves excellent results, the method for processing the input data is poorly described, and no information is provided as to how the temporal domain is used.

In \cite{GraphLSTM}, the authors used a sequence of graphs to model the time evolution of the network traffic, in which nodes are the hosts (in the form of IP addresses) of the network, and edges are the packets exchanged by hosts. After generating these graphs, time series are extracted from each node by computing a set of graph-based features (e.g., \textit{in/out-degree}, or \textit{in/out-neighbors}). Then, such time series are divided in five overlapped time intervals, which are used to train a LSTM to detect malicious (i.e., anomalous) hosts. We observe that the approach in~\cite{GraphLSTM} presents a heavy burden in terms of computational resources for its feature extraction; furthermore, dividing a time series into fixed intervals does not allow to infer long-lasting temporal patterns--despite obtaining encouraging detection results on the considered testbed. Finally, this approach focuses on analyzing PCAP data, and not on NetFlows.

The authors of~\cite{DLFlowSummary} propose an approach similar to~\cite{GraphLSTM}, but assume NetFlows as input. A single graph was extracted from the dataset, where nodes represent hosts (as IP addresses), while edges represent the NetFlows exchanged by the hosts; then, a statistical summary of the NetFlows between each pair of nodes is created. Such statistical summary is integrated with temporal features extracted from the sequence of connection states between each pair of nodes. Then, both the statistical summary and the temporal features are used to make the final inference throughout a FNN. What makes~\cite{DLFlowSummary} intriguing is that there is an actual benefit from using temporal features extracted from the connection states of NetFlows.
However, it also demonstrated that using only the temporal features extracted from the connection states was not enough to achieve good detection.

To summarize, the major drawbacks of these approaches is their specificity to packet level analyses. Inferring spatial features from NetFlow data may still give a benefit, but the overall detection process must take into account the complex NetFlow semantic. Finally, none of these works prove the existence of malicious temporal patterns in NetFlows that are usable for ML-NIDS.

\subsection{Manual extraction}
\label{sec:manual}

These approaches use the features as they are provided in the input data; some preprocessing may occur, but there is no automatic feature extraction mechanism, that is, the features are not produced by any ML method.

In~\cite{AnomalyRNN}, the authors use a bidirectional LSTM model to predict whether a communication was atypical or not. Specifically, they leverage Natural Language Processing by transforming NetFlows into `tokens', where a token is defined as either a \textit{network port} or a \textit{byte-port} tuple. Then, these tokens were grouped by IP-pairs within hourly bins to form time-ordered sequences. This allowed the LSTM to learn a model of the benign traffic, which is used to detect anomalous communications. We observe, however, that the NetFlow semantic was used only to extract information on the \textit{byte-port} tuple, which is a limitation in large networks with high diversity in the traffic. Moreover, having sequences focused on IP-pair and of limited duration does not imply that the learned temporal patterns (if any) are truly effective for NIDS.

In~\cite{AggregationRules}, the authors extend~\cite{AnomalyRNN} by evaluating five different methods for aggregating NetFlow tokens in time-ordered sequences. They showed that different aggregation methods do not play an important role in training sequential models, due to negligible performance differences.
While this observation might be true in their specific context, it is not applicable for our case. Indeed, we want our ML-NIDS to learn the temporal pattern of attacks, thus splitting the corresponding attack NetFlows is not viable: by doing so, we would break the temporal patterns (if they exist).

In~\cite{BotnetVAE}, the authors propose a Recurrent Variational Autoencoder to detect botnet connections through anomaly detection. Their model is trained on certain botnet types and tested on unseen ones, showing good detection performance and demonstrating generalization capabilities. We point out that the generalization capabilities of some ML methods (and, specifically, of RNN) are an appreciable characteristic. However, the sequences used in~\cite{BotnetVAE} are created by aggregating NetFlows on their \textit{source IP}, and then each sequence is reduced by summarizing NetFlows, therefore inducing loss of data which may break any temporal pattern.

In~\cite{BotnetRNN} a behavioral LSTM is devised to counter botnet attacks by analyzing the long-term traffic characteristics. Here, NetFlows are first aggregated into sequences according to a 4-tuple \textit{(source IP, destination IP, destination part, protocol)}; then, each NetFlow is vectorized with the behavioural model on the basis of three features: \textit{size, duration, periodicity}.
Although this work evaluates the prominent issues in training LSTM (e.g., data imbalance), it manually cherry picked the best features for detection. We do not make such assumption in this paper.

The LSTM is used also in~\cite{NIDSLSTMEmbedding} to detect and classify malicious network packets. The authors evaluate the semantic of categorical features through embedding layers. Their results show that such layers do not improve the detection, and that the binary classification of LSTM achieves optimal performance ($99.75\%$ F1-score). However, the LSTM is trained on sequences of fixed length, and no information on how to determine such length is provided. As in~\cite{BotnetRNN}, this paper cherry-picks the optimal thresholds that ensure best results on their specific testbed, with low practical value.

To summarize, these works have little realistic usage, and do not allow to assess the effectiveness of the extracted temporal patterns.

\section{Proposed Solution}
\label{sec:methodology}

Our goal is to investigate the existence of temporal patterns in malicious NetFlows, and then explore the benefits of using these patterns in detecting malicious activities via machine learning.
Hence, our first objective is the extraction of such temporal patterns. 
To this purpose, we propose an original method based on expert knowledge and network data analytics.
Then, we need to verify if the extracted patterns are useful for detection purposes. 

As explained in Section~\ref{sec:related}, many state-of-the-art approaches for extraction of temporal patterns have significant limitations. Thus, we propose an original method that aims at mitigating such deficiencies. The intention is to give more guarantees that the underlying temporal characteristics of different attacks are truly captured by the generated sequences, and are hence usable for efficient detection.
In the remainder of this work, we will use the term `scenario' to denote a complete `attack scenario', i.e., a sequence of NetFlows that describes a specific attack scenario.

We first describe the threat model, the characteristics of the network environment and the assumptions required by our method (Section~\ref{ssec:threat}). Then we present our proposed method for creating scenarios (Section~\ref{ssec:solution} and outline the characteristics of the sequential ML model (Section~\ref{ssec:LSTM}).

\subsection{Threat Model and Assumptions}
\label{ssec:threat}
This paper makes the following assumptions about the attacker's side, the considered network environment, and the input data required for creating the attack scenarios.

\textit{Attacker.} 
We assume an attacker with the goal of either stealing sensitive data from, or disrupting the services provided or used by, the targeted organization. As is common in NIDS scenarios, the attacker is assumed to have established a foothold within the targeted network by compromising one (or more) of its most vulnerable hosts. The attacker does not have any information about the defensive mechanisms monitoring the organization; for example, the attacker does not know the datasets used to train ML-based NIDS. The attacker can interact with the network services accessible from within the organization network, but they do not have direct access to the NetFlow-exporter nor to any defensive system.

\textit{Environment.} 
We assume the typical network environment spanning from dozens to hundreds of hosts that perform multiple network activities. The traffic generated by these hosts is first captured as PCAP data, and then transformed into NetFlows by means of any NetFlow-generation software\footnote{Exemplary NetFlow software: \url{https://qosient.com/argus/}.}. 
These extracted NetFlows are then analyzed by the NIDS, possibly after some preprocessing operations. We assume that the traffic generated by the most critical servers of the organization (to which the attacker has no access) is analyzed by dedicated mechanism. 
    
\textit{Input Data.} 
Our method requires a collection of NetFlows that are labelled according to the malicious network activities that they represent (or as `benign' if they are not related to illegitimate actions). For each NetFlow, our method requires the following pieces of information: the \textit{timestamp}, the \textit{Source/Destination ports}, the \textit{protocol}, the \textit{duration}, the \textit{direction} of the connection with respects to the network environment (i.e., inbound, outbound, or bidirectional), and optionally the specific \textit{attack type} denoted by the label. Any additional piece of information is not required, but -- if available -- it can be freely added to the list of considered features to provide more fine-grained results.

In summary, our proposed method makes the assumptions typical of NIDS scenarios, and only requires the essential pieces of information obtainable from any NetFlow-generation software. The availability of labelled data is also a standard assumption in machine learning tasks.

\subsection{Extraction of Attack Scenarios}
\label{ssec:solution}
Our method uses expert knowledge that can be applied to existing labelled NetFlow data to extract the attack scenarios.

The fundamental idea is to use the available information on a given set of NetFlow data to extract these scenarios. Such information can be readily acquired from the labels or the documentation of a dataset; or after exploratory data-analytics procedures. Our intuition is that if a specific attack presents temporal patterns, then such patterns should not depend on the benign traffic. Hence, we propose to \textit{isolate the specific attack types} and, for each type, focus on the time-interval containing the malicious samples.

However, many obstacles must be overcome to increase the likelihood that the extracted scenarios are effective for detection.

A well-known issue in the application of ML-based approaches for NIDS is the strong unbalancing between the normal `benign' traffic, and the malicious traffic. Indeed, contrarily to other domains where ML is applied, in cybersecurity the illegitimate samples represent rare events (e.g.,~\cite{MalHosts,chuang2019applying}) which are several orders of magnitude inferior to the normal events occurring in a network. A strong data imbalance affects all types of ML applications; however, our time-sensitive context further aggravates this problem, because \textit{finding temporal patterns in the minority class} is more difficult.

Therefore, in designing our proposed methodology for extracting the attack scenarios, we must first mitigate the unbalancing in the data. By doing so, we can improve the quality (in terms of detection efficacy) of the extracted temporal patterns.
To mitigate the unbalancing problem, we leverage two observations:
\begin{enumerate}
    \item the majority of the benign traffic that skews the distribution of a dataset presents similar characteristics, i.e., same protocols or same (service) ports\footnote{This is the assumption underlying anomaly-based NIDS, where a representative model of the benign network traffic is constructed through statistical analyses~\cite{survey2019, mishra2018detailed}.
    }, and can be partially discarded to increase the `weight' of the malicious patterns;
    \item some attacks exhibit periodic behaviours (also referred to as \textit{beaconing}~\cite{MalHosts}), and are executed in time intervals that, if found, can be used to ``split'' the scenario.
\end{enumerate}

By combining these two observations, we propose to identify one or more separate scenarios for each attack type; such scenarios contain only those samples (benign and malicious) generated at the precise time intervals.
Hence, we focus on pinpointing one or more time intervals for each attack, and then create a dedicated scenario for each time interval by including only those NetFlow samples occurring within the corresponding intervals. The remaining traffic samples are then discarded. 
\begin{figure*}[!htbp]
\centerline{\includegraphics[width=6in]{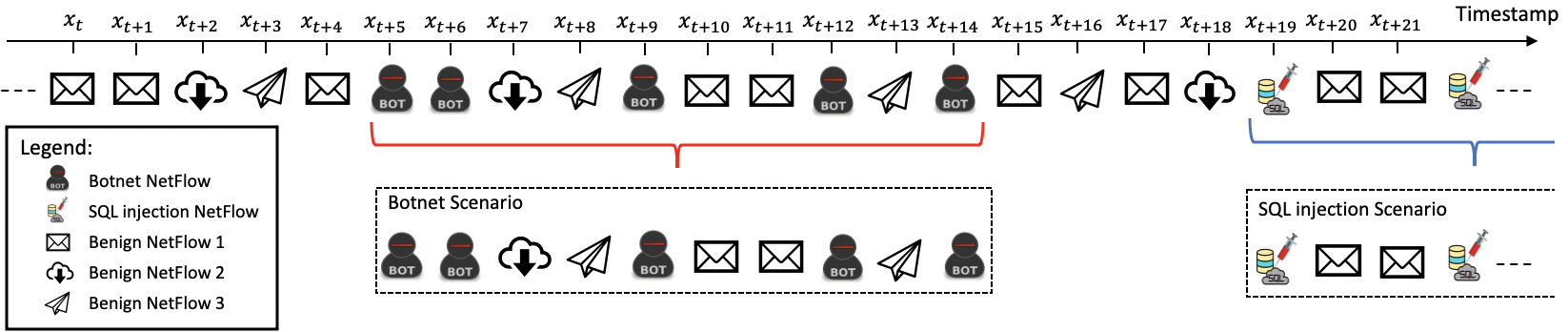}}
\caption{Example of attack scenarios creation. Any NetFlow not included in the attack scenarios is discarded.}
\label{fig:SequenceCreation}
\end{figure*}

Let us illustrate our idea with a concrete example, graphically depicted in Figure~\ref{fig:SequenceCreation}. Here, we assume that there exists a ``Botnet Scenario'' that starts from time $t+5$ (denoted with the NetFlow $x_{t+5}$), and ends at time $t+14$ (denoted with the NetFlow $x_{t+14}$). Our approach retains all NetFlows occurring within such time interval (spanning over 10 time-steps), regardless of their nature (benign or malicious).
Specifically, the extracted ``Botnet Scenario'' will include two botnet NetFlows ($x_{t+5}$, $x_{t+6}$) followed by two benign NetFlows ($x_{t+7}$, $x_{t+8}$), and then by, in order, another botnet NetFlow ($x_{t+9}$), two benign NetFlows ($x_{t+10}$, $x_{t+11}$), a botnet NetFlow ($x_{t+12}$), a benign NetFlow ($x_{t+13}$), and the final botnet NetFlow ($x_{t+14}$)
The NetFlows not included in such scenario (e.g., those from $x_t$ to $x_{t+4}$ and from $x_{t+15}$ to $x_{t+18}$) are discarded. Then, at time $t+19$, the ``SQL Injection Scenario'' begins.

The time intervals of an attack can be identified by analyzing the temporal distribution of malicious NetFlows (e.g., through the use of autocorrelation~\cite{shirani2015method}). However, such intervals can be obtained also from a dataset documentation (e.g.,~\cite{CICIDS2017}).

When identifying the time-intervals of the attack scenarios, it is important to avoid `removing' excessive amounts of benign samples: such occurrences may lead to oversimplified attack scenarios that are not sufficiently representative for detection purposes. If the extracted scenarios do not include at least some benign samples, we recommend increasing the length of the time-intervals to include more benign activities; alternatively, it is even possible to create dedicated `benign' scenarios.
We anticipate that, in our evaluation, we demonstrate that this latter technique can be used without affecting the learning phase of the sequential models.

Nonetheless, there may also exist some circumstances that lead to scenarios with an overabundance of benign samples (e.g., long duration attacks such as DDoS). Completely solving this issue is unfeasible. Hence, to mitigate this problem, we recommend to configure the \textit{loss function} (i.e., the weights) of the ML model to assign a heavier penalty to misclassifications of malicious scenarios.

Let us explain the motivation behind such design choices. The ML literature proposes many techniques for dealing with imbalanced datasets. Some prominent examples involve: under-sampling the majority class or over-sampling the minority class~\cite{cieslak2006combating}; Synthetic Minority Over-sampling Technique (SMOTE) (e.g.,~\cite{SMOTE}); and weighing the loss function.
However, with the exception of weighing of the loss function (which is the approach adopted in the proposed method), all these techniques artificially modify the input data and cannot be readily applied. Indeed, manipulating the network data with under/over-sampling may lead to biased and overfit models that learn unrealistic patterns, which may present good results on the `modified' source dataset, but with impractical performance in realistic applications.
We argue that these techniques are effective for training successful ML models that do not take into account temporal relationships. However, in our time-sensitive context, we have to ensure that the generation of synthetic data - or the removal of real data - does not interfere with the actual temporal patterns that denote the network activities.

Our method is among the first attempts in this domain, and presents much room for improvement.
The major advantage of the proposed method, that is, creating specific attack scenarios, is that it should improve the quality of the learning phase by reducing the noise caused by the benign traffic. However, we stress that \textit{there is no guarantee} that the extracted scenarios capture meaningful temporal relationships that are truly useful for detection purposes. This is why we must resort on ML methods to validate if the proposed method is effective or not.

\subsection{Sequential ML for NIDS}
\label{ssec:LSTM}

To assess the quality of the temporal patterns hopefully contained in the attack scenarios (described in Section~\ref{ssec:solution}), we employ the de-facto sequence model for practical applications, \emph{gated RNN}~\cite{GRU}. Among the possible architectures of gated RNNs, we chose the Long Short-Term Memory (LSTM), due to its successful results achieved in literature (e.g.,~\cite{yeom2021source, mirza2018computer}).

The successful principle at the base of the LSTM is an internal and context-aware self-loop, which dynamically adjusts its time scale on the basis of the sequence provided as input~\cite{sainath2015convolutional}. 
The ability of automatically adjusting the time scale makes LSTM a suitable candidate for our objective.
Indeed, we stress that our method does not know whether temporal patterns (among malicious NetFlow data) exist or not. Moreover, even if such patterns exist, the method does not know when they occur and how long they last. Hence, we require a technique that is able to learn both long-term and short-term dependencies in the input scenarios. As an additional benefit, LSTM are among the best models for dealing with the so called \textit{vanishing gradient} problem~\cite{fei2018bidirectional}, which is a major obstacle when training RNN methods.

\begin{figure}[!htbp]
    \includegraphics[width=0.95\columnwidth]{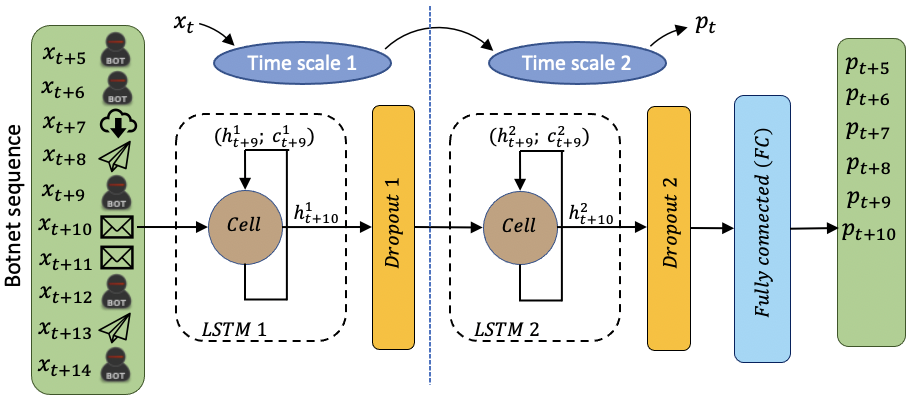}
    \caption{Exemplary architecture of a 2-stacked LSTM.}
    \label{fig:LSTM}
\end{figure}

We propose the use of an unidirectional $n$-stacked version of the LSTM. We provide a schematic depiction of such architecture (in a 2-stacked version) in Figure~\ref{fig:LSTM}, which reports the exemplary case of the Botnet attack scenario shown in Figure~\ref{fig:SequenceCreation}.
As described in~\cite{StackedRNN}, having a stacked version of a RNN helps the model to look at the input data from different time-scales. This is possible since the input of each layer is the hidden state of the previous layer, with the sole exception of the very first layer which receives the actual input $x_t$.

As shown in Figure~\ref{fig:LSTM}, in a $n$-stacked version of a LSTM, the input at time-step $t$ is fed into the first layer, and it is then used (alongside the hidden state of the previous time-step $h_{t-1}^{1}$) to produce the \textit{Cell} state $c_t^1$ (we use the superscript for identifying each layer).
Then, for all the upper $n-1$ layers, the input at time-step $t$ is the hidden state of the previous layer, that is, for a layer $i \ge 2$, the input at time-step $t$ is $h_{t}^{i-1}$.
This mechanism allows to bring the actual input into different time scales before making the final inference.

A common problem in any ML approach is the risk of realizing an overfit model. Depending on the input data and the chosen ML algorithm, different techniques can be used to remove (or mitigate) such issues (e.g.,~\cite{jindal2016learning}). Similarly, the activation function of the LSTM layers can be chosen arbitrarily: common solutions are, e.g., the \textit{sigmoid} or the rectified linear unit (ReLU). 

The model is used in a many-to-many fashion, that is, each NetFlow in a scenario is classified as either benign or malicious. This final classification is achieved through a single fully-connected neuronal layer placed after the last LSTM layer.

\section{Experimental Settings}
\label{sec:experiments}

We experimentally evaluate the proposed methodology. Recall the twofold goal of our paper: investigating the existence of malicious temporal patterns in NetFlows; and verifying whether sequential learning models really learn and benefit from these patterns. To this purpose, we need to first devise a sequential learning model using some input data, and then compare its performance against a baseline `static' ML model (using the same input data), but that does not consider any temporal relationship.

We describe the experimental settings to highlight the realistic value of our testbed. We present the chosen datasets in Section~\ref{ssec:datasets}, the implementation of the proposed method and the realization of the final detectors in Section~\ref{ssec:implementation}.

\subsection{Datasets and Preprocessing}
\label{ssec:datasets}

We base our evaluation on two well-known and labelled datasets of enterprise network traffic that include different types of attacks and that are appreciated in the NIDS domain: the CTU13~\cite{CTU13} and the CICIDS2017~\cite{CICIDS2017}. We present the motivations of our choice below.

\textit{CTU13.} 
This dataset is extensively used in literature (see Table~\ref{tab:works}) and contains thirteen collections of real network traffic data (in the form of NetFlows) mixed with traffic generated by different \textit{botnet} families.
We choose the CTU13 for two reasons that make it as a valid benchmark for realistic evaluations. First, in each collection, a specific botnet family performs multiple malicious actions (e.g., Port Scan, Click Fraud, DDoS); hence, each collection embeds a large variance of malicious events--and potentially many different temporal patterns. Second, the volume of benign traffic with respect to the malicious one is several orders of magnitude greater, making the dataset highly imbalance.

We perform a preliminary analysis of the CTU13. We discover that the majority of the benign traffic in CTU13 is produced by 20 internal hosts and involve few services. Hence, we exclude the NetFlows (which are all benign) of these hosts to partially mitigate the unbalancing problem. We recall that our focus is investigating the existence of temporal patterns among malicious NetFlows, and removing such samples does not interfere with our methodology.
Furthermore, these hosts can be seen as more critical to the organization, and in real networks they would be protected by dedicated defensive mechanisms, therefore aligning with the assumptions made in our threat model (see Section~\ref{ssec:threat}).
As successfully done in related NIDS literature (e.g.,~\cite{apruzzese2020deep}), we select the following NetFlow features\footnote{Note that our chosen features extend the basic essential features required by our method (see Section~\ref{ssec:solution}).}: duration, source/destination ports, direction, protocol, total packets, packets per second, total amount of bytes, amount of source bytes, ratio of bytes per second, ratio of bytes per packet.

\textit{CICIDS2017.}
This benchmark dataset contains synthetic benign traffic and common types of attacks, with the peculiarity that different types of attacks are carried out in distinct time intervals.
We choose the CICIDS2017 because, with respect to the CTU13, it is more recent, it has a larger variety of attacks and it is less class imbalance. Moreover, its documentation specifies the time interval of each type of attack, which we leverage for our method.

We conduct an exploratory analysis of the CICIDS2017, where we determine the features for our experiments: duration, protocol, source/destination IP (as either \textit{internal} or \textit{external} to derive the direction of the connection), total packets, source/destination ports, total amount of bytes, ratio of packets per second, ratio of bytes per packet, minimum/maximum/average inter arrival time. This feature set differs from the one adopted in the CTU13, but it includes the features required by our approach.

\begin{table*}[t!]
    \caption{The distribution of benign and malicious NetFlows in the datasets before and after the application of the method described in Section \ref{ssec:solution}. In the table, \#B stands for number of benign and \#M for number of malicious.}
    \arrayrulecolor{white}
    \resizebox{\textwidth}{2.8cm}{
    \begin{tabular}{@{}c|c@{}}
    \begin{tabular}{c|cc|ccccc}
        \headerline
        \multicolumn{8}{c}{\header{CICIDS 2017}} \\
        \hline
        \headerline
        \multicolumn{3}{c|}{\header{NetFlows in the dataset}} & \multicolumn{5}{c}{\header{NetFlows in processed scenarios}} \\
        \headerline
        {\headerb{File}} & 
        {\headerb{\# NetFlows}} & 
        {\headerb{\# Mal}} & 
        {\headerb{Scenario}} & 
        {\headerb{\#B train}} & 
        {\headerb{\#M train}} & 
        {\headerb{\#B test}} & 
        {\headerb{\#M test}} \\
        \lightrow 
        \darkcell{} & \darkcell{} & \darkcell{} & 
        \textit{Benign1} & \ncell{191235} & \ncell{0} & \ncell{81958} & \ncell{0} \\
        \darkrow 
        \multirow{-2}{*}{\darkcell{ Day 1}} & 
        \multirow{-2}{*}{\darkcell{ {516243}}} & 
        \multirow{-2}{*}{\darkcell{ 0}} & 
        \textit{Benign2} & \ncell{170135} & \ncell{0} & \ncell{72915} & \ncell{0} \\
        \hline
        \lightrow 
        \darkcell{} & \darkcell{} & \darkcell{} &
        \textit{Botnet1} & \ncell{105288} & \ncell{1460} & \ncell{45538} & \ncell{212} \\
        \darkrow 
        \darkcell{} & \darkcell{} & \darkcell{} &
        \textit{Botnet2} & \ncell{9607} & \ncell{107} & \ncell{4019} & \ncell{145}\\
        \lightrow 
        \darkcell{} & \darkcell{} & \darkcell{} & 
        \textit{Port Scan} & \ncell{ 88585} & \ncell{ 117607} & \ncell{ 47173} & \ncell{ 41196} \\
        \darkrow 
        \multirow{-4}{*}{\darkcell{Day 2}} &
        \multirow{-4}{*}{\darkcell{\ncell{687858}}} &
        \multirow{-4}{*}{\darkcell{\ncell{284758}}} &
        \textit{DDoS} & \ncell{ 42946} & \ncell{ 101443} & \ncell{ 39294} & \ncell{ 22588} \\
        \hline
        \lightrow 
        \darkcell{} & \darkcell{} & \darkcell{} & 
        \textit{DoS GoldenEye} & \ncell{ 10202} & \ncell{ 7690} & \ncell{ 24456} & \ncell{ 2384} \\
        \darkrow 
        \darkcell{} & \darkcell{} & \darkcell{} & 
        \textit{DoS Hulk} & \ncell{16802} & \ncell{153596} & \ncell{14358} & \ncell{58670} \\
        \lightrow 
        \darkcell{} & \darkcell{} & \darkcell{} & 
        \textit{DoS slowhttptest} & \ncell{19655} & \ncell{3974} & \ncell{14227} & \ncell{1527}\\
        \darkrow 
        \multirow{-4}{*}{\darkcell{ Day 3}} & 
        \multirow{-4}{*}{\darkcell{ \ncell{663612}}} & 
        \multirow{-4}{*}{\darkcell{ \ncell{233635}}} & 
        \textit{DoS slowloris} & \ncell{54111} & \ncell{3878} & \ncell{8323} & \ncell{1911} \\
        \hline
        \lightrow 
        \darkcell{} & \darkcell{} & \darkcell{} & 
        \textit{FTP-Patator} & \ncell{73355} & \ncell{5027} & \ncell{23262} & \ncell{2866} \\
        \darkrow 
        \multirow{-2}{*}{\darkcell{Day 4}} & \multirow{-2}{*}{\darkcell{\ncell{436088}}} & \multirow{-2}{*}{\darkcell{\ncell{13782}}}  & 
        \textit{SSH-Patator} & \ncell{55133} & \ncell{3522} & \ncell{22772} & \ncell{2366} \\
        \hline
        \lightrow 
        \darkcell{} & \darkcell{} & \darkcell{} &
        \textit{Web Brute Force} & \ncell{43062} & \ncell{867} & \ncell{18188} & \ncell{640} \\
        \darkrow 
        \multirow{-2}{*}{\darkcell Day 5} &
        \multirow{-2}{*}{\darkcell{\ncell{448382}}} &
        \multirow{-2}{*}{\darkcell{\ncell{2214}}} &
        \textit{Web XSS} & \ncell{15984} & \ncell{417} & \ncell{6794} & \ncell{235} \\
        \greenrow
        Summary & \ncell{2752183} & \ncell{534389} & 
        Summary & \ncell{896100} & \ncell{399588} & \ncell{423277} & \ncell{134740}
    \end{tabular}
    &
    \begin{tabular}{c|cc|ccccc}
        \headerline
        \multicolumn{8}{c}{\header{CTU 13}} \\
        \hline
        \headerline
        \multicolumn{3}{c|}{\header{NetFlows in the dataset}} & 
        \multicolumn{5}{c}{\header{NetFlows in processed scenarios}} \\
        \headerline
        {\headerb{File}} & 
        {\headerb{\# NetFlows}} & 
        {\headerb{\# Mal}} & 
        {\headerb{Scenario}} & 
        {\headerb{\#B train}} & 
        {\headerb{\#M train}} & 
        {\headerb{\#B test}} & 
        {\headerb{\#M test}} \\
        \lightrow 
        Menti & \ncell{558919} & \ncell{4630} & 
        \textit{Menti} & \ncell{62306} & \ncell{2986} & \ncell{26369} & \ncell{1614} \\
        \darkrow 
        Murlo & \ncell{2954230} & \ncell{6126} & 
        \textit{Murlo} & \ncell{273249} & \ncell{3641} & \ncell{116251} & \ncell{2417} \\
        \lightrow 
        Neris1 & \ncell{2824636} & \ncell{40959} & 
        \textit{Neris1} & \ncell{156026} & \ncell{24280} & \ncell{68013} & \ncell{9261} \\
        \darkrow 
        Neris2 & \ncell{1808122} & \ncell{20941} & 
        \textit{Neris2} & \ncell{99614} & \ncell{16821} & \ncell{46569} & \ncell{3333} \\
        \lightrow 
        Neris3 & \ncell{2753884} & \ncell{184979} & 
        \textit{Neris3} & \ncell{86236} & \ncell{51890} & \ncell{33908} & \ncell{25290} \\
        \darkrow 
        Nsisay & \ncell{325471} & \ncell{2168} & 
        \textit{Nsisay} & \ncell{32228} & \ncell{1602} & \ncell{14140} & \ncell{359} \\
        \lightrow 
        Rbot1 & \ncell{4710638} & \ncell{26822} & 
        \textit{Rbot1} & \ncell{317008} & \ncell{12451} & \ncell{126869} & \ncell{14329} \\
        \darkrow 
        Rbot2 & \ncell{1121076} & \ncell{1768} & 
        \textit{Rbot2} & \ncell{53924} & \ncell{825} & \ncell{22288} & \ncell{1177} \\
        \lightrow 
        Rbot3 & \ncell{1309791} & \ncell{106352} & 
        \textit{Rbot3} & \ncell{112684} & \ncell{74743} & \ncell{48939} & \ncell{31388} \\
        \darkrow 
        Rbot4 & \ncell{107251} & \ncell{8164} & 
        \textit{Rbot4} & \ncell{5598} & \ncell{5070} & \ncell{1500} & \ncell{3073} \\
        \lightrow 
        Virut1 & \ncell{129832} & \ncell{901} & 
        \textit{Virut1} & \ncell{8331} & \ncell{549} & \ncell{3501} & \ncell{306} \\
        \darkrow 
        Virut2 & \ncell{1925149} & \ncell{39993} & 
        \textit{Virut2} & \ncell{141886} & \ncell{23225} & \ncell{62048} & \ncell{8714} \\
        \greenrow
        Summary & \ncell{20528999} & \ncell{443803} & 
        Summary & \ncell{1349090} & \ncell{218083} & \ncell{570395} & \ncell{101261} \\
    \end{tabular}
    \end{tabular}
    }
    \label{tab:datasets}
\end{table*}

\subsection{Implementation and Detectors}
\label{ssec:implementation}
After preprocessing the datasets, the next step is the extraction of the attack scenarios with the proposed method (see Section \ref{ssec:solution}). Its application produces a set of attack scenarios, which we summarize in Table~\ref{tab:datasets}. Let us describe the table by focusing on its leftmost part: the dataset is the CICIDS2017 for which precise information on the time interval of the attacks is available. We use such information to create the scenarios.
For instance, \emph{Day 2} is split into 4 scenarios (\textit{Botnet1, Botnet2, Port Scan}, and \textit{DDoS}). We observe that our method reduces the amount of benign NetFlows: in \emph{Day 2} they decrease from $687$k to $382$k. Furthermore, instead of dropping \textit{Day 1}, we use its benign traffic to create two benign scenarios to demonstrate that their use can yield better performance.

To assess the quality of the extracted patterns for detection purposes, we devise a baseline detector that relies only on the ``static'' features of NetFlows for the final inference, that is, it does not take into account any temporal relationship. The rationale is that, if the baseline `static' detector and the proposed `sequential' detector have comparable overall performance, but their specific classification patterns differ, we can prove that temporal patterns among malicious NetFlows indeed exist in our scenarios.

For the sequential detector, we use a 2-stacked LSTM. To mitigate overfitting, the LSTM leverages the \textit{dropout} technique to yield a smoother model that is less sensitive to small variations in the (unseen) input data. Moreover, we use the \textit{sigmoid} as activation function. The LSTM has the same structure as in Figure~\ref{fig:LSTM}.

\begin{figure}[!htbp]
    \includegraphics[width=0.9\columnwidth]{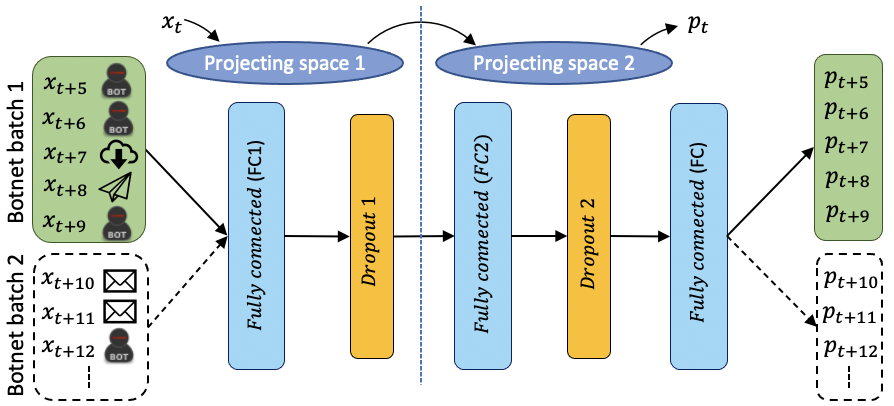}
    \caption{FNN architecture.}
    \label{fig:FNN}
\end{figure}

For the static detector, we consider a Feedforward Neural Network (FNN). To avoid bias and favor a fair comparison, we build the FNN with the same number of hidden layers of the LSTM; we apply the dropout mechanism after each layer, and we use \textit{ReLU} as activation function.
We report such FNN architecture in Figure~\ref{fig:FNN}, which can be directly compared to the LSTM in Figure~\ref{fig:LSTM}.

We observe that both the LSTM and the FNN project the NetFlows in two latent spaces, the blue ellipsis.
While the LSTM projects sequences of NetFlows in two distinct and subsequent time scales, the FNN separately transforms each NetFlow into two subsequent projection spaces.
The models should have a similar expressiveness (i.e., optimization functions) which is beneficial for a fair comparison.

We focus on binary classification, that is, each sample can be predicted to be either benign or malicious. We consider two datasets, so we devise a total of four models, i.e., one FNN and one LSTM per dataset. 
We report in Table~\ref{tab:datasets}, the partitioning of the training and testing sets. We use the first $70\%$ of each scenario for training and the remaining for testing (to avoid temporal bias~\cite{pendlebury2019tesseract}); the only exceptions are some scenarios of CICIDS2017, where we must resort to different splits to ensure at that at least $70\%$ of the malicious NetFlows are put in the training set. We remark that both the LSTM and FNN are trained on the \textit{same} training data, and tested on the \textit{same} test data. Note also that we keep the \textit{Benign1-2} scenarios of CICIDS2017 only for training the models.

\section{Evaluation and Results}
\label{sec:results}

The goal of the experiments is verifying the existence and effectiveness of the temporal patterns extracted with the proposed method; we do not want to propose a superior technique. Hence, we train\footnote{We report more experimental details in Appendix~\ref{app:hyperparameters}.} the LSTM and FNN on each dataset (as described in Section~\ref{ssec:implementation}), and we perform two types of experiments, which we briefly outline.

\textit{Direct comparison.}
Here, the focus is studying the classification patterns learned by the two models (LSTM and FNN). We do so by analyzing the results of both models on the same testing set. The idea is that different performance implies different learned patterns used for detection. We present and discuss the results of these experiments in Section~\ref{ssec:performance}.

\textit{Ensemble comparison.}
To truly verify if the two models (FNN and LSTM) learn different patterns, we create a third `ensemble' model that includes both models. More specifically, we join the FNN with the LSTM model through the \logicalor{logical or}. Then, we compare this ensemble (denoted as FNN+LSTM) with the stand-alone LSTM model, and test them on the same test data. The intuition is as follows: if we combine the LSTM with the FNN and the performance matches the performance of the stand-alone LSTM, then it will mean that the FNN and LSTM learn the same detection patterns (and vice versa). These experiments are discussed in Section~\ref{ssec:ensemble}.

\subsection{Initial Assessment (FNN vs LSTM)}
\label{ssec:performance}

We first compare the performance of the sequential LSTM against the static FNN. We evaluate the two classifiers on every scenario of each dataset. We report the results in Table \ref{tab:LSTM_FNNResults}, showing the performance metrics of interest: the F1-score; and the complete confusion matrix, where \textit{fp} (\textit{fn}) stands for false positives (negatives), and \textit{tp} (\textit{tn}) stands for true positives (negatives). We consider a positive as a malicious sample.
We split the classification results for each scenario to allow a more fine-grained comparison of the two approaches. The green rows report the aggregated results on the corresponding dataset.

\begin{table*}[t!]
    \caption{Results of the LSTM and FNN on the two datasets.}
    \arrayrulecolor{white}
    \resizebox{.85\textwidth}{4.5cm}{
    \begin{tabular}{@{}c|c@{}}
    \begin{tabular}{c||c|ccccc}
        \headerline
        {\cellcolor{white}} & \multicolumn{6}{c}{\header{CICIDS 2017}} \\
        \hline
        \headerline
        {\cellcolor{white}} & 
        {\headerb{Scenario}} & 
        {\headerb{F1-score \%}} & 
        {\headerb{tp}} & 
        {\headerb{fn}} & 
        {\headerb{tn}} & 
        {\headerb{fp}} \\
        \yellowrow
        {\header{}} & \textit{Botnet1} & \ncell{81.755} & \ncell{177} & \ncell{35} & \ncell{45494} & \ncell{44} \\
        \yellowrow
        {\header{}} & \textit{Botnet2} & \ncell{87.542} & \ncell{130} & \ncell{15} & \ncell{3997} & \ncell{22} \\
        \lightrow 
        {\header{}} & \textit{Port Scan} & \ncell{99.558} & \ncell{41179} & \ncell{17} & \ncell{46824} & \ncell{349} \\
        \darkrow 
        {\header{}} & \textit{DDoS} & \ncell{99.572} & \ncell{22588} & \ncell{0} & \ncell{39100} & \ncell{194} \\
        \lightrow 
        {\header{}} & \textit{DoS GoldenEye} & \ncell{99.707} & \ncell{2384} & \ncell{0} & \ncell{24442} & \ncell{14} \\
        \darkrow 
        {\header{}} & \textit{DoS Hulk} & \ncell{99.955} & \ncell{58653} & \ncell{17} & \ncell{14322} & \ncell{36} \\
        \lightrow 
        {\header{}} & \textit{DoS slowhttptest} & \ncell{98.444} & \ncell{1487} & \ncell{40} & \ncell{14220} & \ncell{7} \\
        \darkrow 
        {\header{}} & \textit{DoS slowloris} & \ncell{99.843} & \ncell{1910} & \ncell{1} & \ncell{8318} & \ncell{5} \\
        \lightrow 
        {\header{}} & \textit{FTP-Patator} & \ncell{99.600} & \ncell{2866} & \ncell{0} & \ncell{23239} & \ncell{23} \\
        \darkrow 
        {\header{}} & \textit{SSH-Patator} & \ncell{98.666} & \ncell{2366} & \ncell{0} & \ncell{22708} & \ncell{64} \\
        \lightrow 
        {\header{}} & \textit{Web Brute Force} & \ncell{99.456} & \ncell{640} & \ncell{0} & \ncell{18181} & \ncell{7} \\
        \darkrow 
        {\header{}} & \textit{Web XSS} & \ncell{99.153} & \ncell{234} & \ncell{1} & \ncell{6791} & \ncell{3} \\
        \greenrow
        \multirow{-14}{*}{\rotatebox[origin=c]{90}{\header{Long Short-Term Memory}}}
         & Summary & \ncell{99.669} & \ncell{134614} & \ncell{126} & \ncell{267636} & \ncell{768} \\
    \end{tabular}
    &
    \begin{tabular}{c|ccccc}
        \headerline
        \multicolumn{6}{c}{\header{CTU 13}} \\
        \hline
        \headerline
        {\headerb{Scenario}} & 
        {\headerb{F1-score \%}} & 
        {\headerb{tp}} & 
        {\headerb{fn}} & 
        {\headerb{tn}} & 
        {\headerb{fp}} \\
        \lightrow 
        \textit{Menti} & \ncell{95.153} & \ncell{1600} & \ncell{14} & \ncell{26220} & \ncell{149} \\
        \darkrow 
        \textit{Murlo} & \ncell{95.640} & \ncell{2369} & \ncell{48} & \ncell{116083} & \ncell{168} \\
        \lightrow 
        \textit{Neris1} & \ncell{93.384} & \ncell{8793} & \ncell{468} & \ncell{67235} & \ncell{778} \\
        \darkrow 
        \textit{Neris2} & \ncell{90.524} & \ncell{3162} & \ncell{171} & \ncell{46078} & \ncell{491} \\
        \yellowrow
        \textit{Neris3} & \ncell{95.911} & \ncell{24901} & \ncell{389} & \ncell{32174} & \ncell{1734} \\
        \darkrow 
        \textit{Nsisay} & \ncell{41.445} & \ncell{109} & \ncell{250} & \ncell{14082} & \ncell{58} \\
        \lightrow 
        \textit{Rbot1} & \ncell{98.105} & \ncell{14080} & \ncell{249} & \ncell{126574} & \ncell{295} \\
        \darkrow 
        \textit{Rbot2} & \ncell{94.873} & \ncell{1101} & \ncell{76} & \ncell{22245} & \ncell{43} \\
        \lightrow 
        \textit{Rbot3} & \ncell{99.909} & \ncell{31352} & \ncell{36} & \ncell{48918} & \ncell{21} \\
        \darkrow 
        \textit{Rbot4} & \ncell{100.000} & \ncell{3073} & \ncell{0} & \ncell{1500} & \ncell{0} \\
        \lightrow 
        \textit{Virut1} & \ncell{87.117} & \ncell{284} & \ncell{22} & \ncell{3439} & \ncell{62} \\
        \yellowrow
        \textit{Virut2} & \ncell{80.851} & \ncell{6863} & \ncell{1851} & \ncell{60648} & \ncell{1400} \\
        \greenrow
        Summary & \ncell{95.703} & \ncell{97687} & \ncell{3574} & \ncell{565196} & \ncell{5199} \\
    \end{tabular}
    \\ \hline \hline
    \begin{tabular}{c||c|P{15mm}cccc}
        \yellowrow
        {\header{}} & \textit{Botnet1} & \ncell{93.735} & \ncell{202} & \ncell{10} & \ncell{45521} & \ncell{17} \\
        \yellowrow
        {\header{}} & \textit{Botnet2} & \ncell{81.569} & \ncell{104} & \ncell{41} & \ncell{4013} & \ncell{6} \\
        \lightrow 
        {\header{}} & \textit{Port Scan} & \ncell{99.773} & \ncell{41152} & \ncell{44} & \ncell{47030} & \ncell{143} \\
        \darkrow 
        {\header{}} & \textit{DDoS} & \ncell{99.817} & \ncell{22588} & \ncell{0} & \ncell{39211} & \ncell{83} \\
        \lightrow 
        {\header{}} & \textit{DoS GoldenEye} & \ncell{99.916} & \ncell{2384} & \ncell{0} & \ncell{24452} & \ncell{4} \\
        \darkrow 
        {\header{}} & \textit{DoS Hulk} & \ncell{100.000} & \ncell{58670} & \ncell{0} & \ncell{14358} & \ncell{0} \\
        \lightrow 
        {\header{}} & \textit{DoS slowhttptest} & \ncell{98.303} & \ncell{1477} & \ncell{50} & \ncell{14226} & \ncell{1} \\
        \darkrow 
        {\header{}} & \textit{DoS slowloris} & \ncell{99.974} & \ncell{1911} & \ncell{0} & \ncell{8322} & \ncell{1} \\
        \lightrow 
        {\header{}} & \textit{FTP-Patator} & \ncell{99.930} & \ncell{2866} & \ncell{0} & \ncell{23258} & \ncell{4} \\
        \darkrow 
        {\header{}} & \textit{SSH-Patator} & \ncell{99.705} & \ncell{2366} & \ncell{0} & \ncell{22758} & \ncell{14} \\
        \lightrow 
        {\header{}} & \textit{Web Brute Force} & \ncell{99.844} & \ncell{640} & \ncell{0} & \ncell{18186} & \ncell{2} \\
        \darkrow 
        {\header{}} & \textit{Web XSS} & \ncell{99.788} & \ncell{235} & \ncell{0} & \ncell{6793} & \ncell{1} \\
        \greenrow
        \multirow{-14}{*}{\rotatebox[origin=c]{90}{\header{Feedforward Neural Network}}} & Summary & \ncell{99.844} & \ncell{134595} & \ncell{145} & \ncell{268128} & \ncell{276} \\
    \end{tabular}
    &
    \begin{tabular}{c|P{14mm}cccc}
        \lightrow 
        \textit{Menti} & \ncell{84.960} & \ncell{1610} & \ncell{4} & \ncell{25803} & \ncell{566} \\
        \darkrow 
        \textit{Murlo} & \ncell{63.267} & \ncell{2413} & \ncell{4} & \ncell{113453} & \ncell{2798} \\
        \lightrow 
        \textit{Neris1} & \ncell{85.188} & \ncell{9087} & \ncell{174} & \ncell{65027} & \ncell{2986} \\
        \darkrow 
        \textit{Neris2} & \ncell{84.533} & \ncell{3189} & \ncell{144} & \ncell{45546} & \ncell{1023} \\
        \yellowrow
        \textit{Neris3} & \ncell{97.353} & \ncell{24605} & \ncell{685} & \ncell{33255} & \ncell{653} \\
        \darkrow 
        \textit{Nsisay} & \ncell{44.673} & \ncell{174} & \ncell{185} & \ncell{13894} & \ncell{246} \\
        \lightrow 
        \textit{Rbot1} & \ncell{84.926} & \ncell{14299} & \ncell{30} & \ncell{121823} & \ncell{5046} \\
        \darkrow 
        \textit{Rbot2} & \ncell{79.619} & \ncell{1170} & \ncell{7} & \ncell{21696} & \ncell{592} \\
        \lightrow 
        \textit{Rbot3} & \ncell{98.688} & \ncell{31355} & \ncell{33} & \ncell{48138} & \ncell{801} \\
        \darkrow 
        \textit{Rbot4} & \ncell{99.854} & \ncell{3073} & \ncell{0} & \ncell{1491} & \ncell{9} \\
        \lightrow 
        \textit{Virut1} & \ncell{75.943} & \ncell{292} & \ncell{14} & \ncell{3330} & \ncell{171} \\
        \yellowrow
        \textit{Virut2} & \ncell{87.553} & \ncell{8047} & \ncell{667} & \ncell{60427} & \ncell{1621} \\
        \greenrow
        Summary & \ncell{91.497} & \ncell{99314} & \ncell{1947} & \ncell{553883} & \ncell{16512} \\
    \end{tabular}
    \end{tabular}
    }
    \label{tab:LSTM_FNNResults}
\end{table*}

Let us discuss these results, starting from the CICIDS2017 dataset. By looking at the overall results (green rows), the performance of both LSTM and FNN is similar, and comparable to the state of the art (e.g.,~\cite{DeepHierarchicalNet, STIDM}). However, by inspecting the performance in the individual scenarios, we can observe some interesting phenomena. Consider the two Botnet scenarios of CICIDS2017 (highlighted in yellow): here, the LSTM learns different classification patterns than the FNN, achieving $81\%$ ($87\%$) F1-score against a $93\%$ ($81\%$) in the Botnet1 (Botnet2) scenario. What is remarkable in these two scenarios is the fact that the malicious samples are from the same botnet type. If we turn the attention to the \emph{DoS slowloris} and the \emph{Port Scan} scenarios, the LSTM has almost the same \textit{fn} of the FNN in the \emph{DoS slowloris} scenario, but fewer in the \emph{Port Scan} scenario. These two different types of attacks further highlight that, against some attacks, the LSTM appears to be more effective. The intriguing observation is that a \textit{Port Scan} typically generates a large volume of (malicious) traffic in a short timeframe; on the contrary, the \textit{DoS slowloris} generates a small volume of traffic but in long timeframes. Despite these differences, the LSTM seems to adjust its internal time scale (see Section~\ref{ssec:LSTM}) for achieving superior performance.

We now focus on the CTU13. Here, on average, both models achieve good performance with over $91\%$ F1-score, in-line with the state of the art (e.g.,~\cite{apruzzese2020deep}). The LSTM performs better than the FNN in terms of F1-score, and it exhibits a lower amount of \textit{fp}, but also slightly higher \textit{fn}, making it a more precise detector for practical purposes.
Let us observe the results for the individual scenarios. In the \textit{Nsisay} scenario both models achieve similar poor performance (below $45\%$ F1-score). The most remarkable differences involve the \emph{Neris3} and \emph{Virut2} scenarios (highlighted in yellow). In \emph{Neris3}, the LSTM has a lower number of false-negatives, but it also has a larger number of false-positives; on the other hand, \emph{Virut2} is the only scenario where the FNN outperforms the LSTM, and this is clear from the F1-scores of the models. Indeed, the \textit{Virut2} scenario is the most `problematic' scenario for the LSTM due to the huge \textit{fn}. However, everywhere else the LSTM learns different classification patterns that produce superior detection results.

\vspace{0.5em}
\takeaway{the different performance in the individual scenarios on both datasets may suggest that the LSTM learns different classification patterns than the FNN. Hence, for some specific attacks, the application of the proposed method (and its usage for LSTM-detectors) may yield better NIDS.}

\subsection{Verification (FNN+LSTM vs LSTM)}
\label{ssec:ensemble}

Despite the differences shown in the previous experiments, we still cannot be truly sure that the LSTM and FNN have learned different patterns for their classification\footnote{The objective is verifying if the temporal relationships are truly useful for ML-NIDS.}. As an example, consider the case of \textit{Botnet1} in the CICIDS2017 dataset: the FNN has 10~\textit{fn}, while the LSTM has 35~\textit{fn}. The question is: are these 10 \textit{fn} by the FNN also included in the 35 \textit{fn} by the LSTM? If this is true, then the two models will have learned (very) similar classification patterns; on the other hand, if this is not true, then the two models will have learned (arguably) different classification patterns. 

To this purpose, we join the LSTM model with the FNN model with the \logicalor{logical or} operator to create a new ensemble (FNN+LSTM). More specifically, for a given test-sample as input, the output will be malicious if at least one model of the ensemble makes a malicious classification; conversely, the output will be benign only if both models agree on a benign classification. Such design choice should reduce the number of \textit{fn}. We \textit{do not retrain} the models of the ensemble: we simply input a (test) sample to each trained model, and then join the output with the \logicalor{logical or}.

We test the LSTM+FNN ensemble on both datasets and report the results in Table~\ref{tab:Ensemble}. For each dataset, the three columns of the FNN+LSTM report the F1-score and the number of \textit{fn} and \textit{fp} for each attack scenario; we summarize the overall results in the last row (green row). The two rightmost columns report the difference ($\Delta$) of each performance metric between the FNN+LSTM and the stand-alone LSTM: hence, in the $\pm$fn and $\pm$fp columns, a negative (positive) value means that the ensemble LSTM+FNN is better (worse) than the stand-alone LSTM.

\begin{table*}[t!]
    \caption{Results of the FNN+LSTM ensemble, and comparison ($\Delta$) with the stand-alone LSTM.}
    \arrayrulecolor{white}
    \resizebox{\textwidth}{2.7cm}{
    \begin{tabular}{@{}c|c@{}}
    \begin{tabular}{c|cc|ccc|cc}
        \headerline
        \multicolumn{8}{c}{\header{CICIDS2017}} \\
        \hline
        \headerline
         &
        \multicolumn{2}{c|}{\header{Testing data}} & \multicolumn{3}{c}{\header{FNN+LSTM}} &
        \multicolumn{2}{c}{\header{$\Delta$}} \\
        \headerline
        \multirow{-2}{*}{\header{Scenario}} & 
        {\headerb{\# NetFlows}} & 
        {\headerb{\# Mal}} & 
        {\headerb{F1-score \%}} & 
        {\headerb{fn}} & 
        {\headerb{fp}} & 
        {\headerb{$\pm$fn}} &
        {\headerb{$\pm$fp}} \\
        \yellowrow 
        \textit{Botnet1} & \ncell{45750} & \ncell{212} & \ncell{89.936} & \ncell{2} & \ncell{45} & \ncell{-33} & \ncell{1} \\
        \darkrow 
        \textit{Botnet2} & \ncell{4164} & \ncell{145} & \ncell{91.558} & \ncell{4} & \ncell{22} & \ncell{-11} & \ncell{0} \\
        \lightrow 
        \textit{Port Scan} & \ncell{88369} & \ncell{41196} & \ncell{99.556} & \ncell{9} & \ncell{358} & \ncell{-8} & \ncell{9} \\
        \darkrow 
        \textit{DDoS} & \ncell{61882} & \ncell{22588} & \ncell{99.572} & \ncell{0} & \ncell{194} & \ncell{0} & \ncell{0} \\
        \lightrow 
        \textit{Dos GoldenEye} & \ncell{26840} & \ncell{2384} & \ncell{99.707} & \ncell{0} & \ncell{14} & \ncell{0} & \ncell{0} \\
        \darkrow  
        \textit{DoS Hulk} & \ncell{73028} & \ncell{58670} & \ncell{99.969} & \ncell{0} & \ncell{36} & \ncell{-17} & \ncell{0} \\
        \yellowrow 
        \textit{DoS slowhttptest} & \ncell{15754} & \ncell{1527} & \ncell{99.443} & \ncell{10} & \ncell{7} & \ncell{-30} & \ncell{0} \\
        \darkrow  
        \textit{DoS slowloris} & \ncell{10234} & \ncell{1911} & \ncell{99.869} & \ncell{0} & \ncell{5} & \ncell{-1} & \ncell{0} \\
        \lightrow
        \textit{FTP-Patator} & \ncell{26128} & \ncell{2866} & \ncell{99.600} & \ncell{0} & \ncell{23} & \ncell{0} & \ncell{0} \\
        \darkrow 
        \textit{SSH-Patator} & \ncell{25138} & \ncell{2366} & \ncell{98.666} & \ncell{0} & \ncell{64} & \ncell{0} & \ncell{0} \\
        \lightrow 
        \textit{Web Brute Force} & \ncell{18828} & \ncell{640} & \ncell{99.456} & \ncell{0} & \ncell{7} & \ncell{0} & \ncell{0} \\
        \darkrow 
        \textit{Web XSS} & \ncell{7029} & \ncell{235} & \ncell{99.366} & \ncell{0} & \ncell{3} & \ncell{-1} & \ncell{0} \\
        \greenrow
        Summary & \ncell{403144} & \ncell{134740} & \ncell{99.703} & \ncell{25} & \ncell{778} & \ncell{-101} & \ncell{10}
    \end{tabular}
    &
    \begin{tabular}{c|cc|ccccc}
        \headerline
        \multicolumn{8}{c}{\header{CTU13}} \\
        \hline
        \headerline
         &
        \multicolumn{2}{c|}{\header{Testing data}} & \multicolumn{3}{c}{\header{FNN+LSTM}} &
        \multicolumn{2}{c}{\header{$\Delta$}} \\
        \headerline
        \multirow{-2}{*}{\header{Scenario}} & 
        {\headerb{\# NetFlows}} & 
        {\headerb{\# Mal}} & 
        {\headerb{F1-score \%}} & 
        {\headerb{fn}} & 
        {\headerb{fp}} & 
        {\headerb{$\pm$fn}} &
        {\headerb{$\pm$fp}} \\
        \lightrow 
        \textit{Menti} & \ncell{27983} & \ncell{1614} & \ncell{82.688} & \ncell{2} & \ncell{673} & \ncell{-12} & \ncell{524} \\
        \darkrow 
        \textit{Murlo} & \ncell{118668} & \ncell{2417} & \ncell{62.562} & \ncell{4} & \ncell{2884} & \ncell{-44} & \ncell{2716} \\
        \yellowrow 
        \textit{Neris1} & \ncell{77274} & \ncell{9261} & \ncell{84.099} & \ncell{130} & \ncell{3323} & \ncell{-338} & \ncell{2545} \\
        \darkrow 
        \textit{Neris2} & \ncell{49902} & \ncell{3333} & \ncell{82.624} & \ncell{33} & \ncell{1355} & \ncell{-138} & \ncell{864} \\
        \lightrow 
        \textit{Neris3} & \ncell{59198} & \ncell{25290} & \ncell{95.938} & \ncell{86} & \ncell{2048} & \ncell{-303} & \ncell{314} \\
        \darkrow 
        \textit{Nsisay} & \ncell{14499} & \ncell{359} & \ncell{50.228} & \ncell{139} & \ncell{297} & \ncell{-111} & \ncell{239} \\
        \lightrow 
        \textit{Rbot1} & \ncell{141198} & \ncell{14329} & \ncell{84.654} & \ncell{11} & \ncell{5180} & \ncell{-238} & \ncell{4885} \\
        \darkrow 
        \textit{Rbot2} & \ncell{23465} & \ncell{1177} & \ncell{79.268} & \ncell{7} & \ncell{605} & \ncell{-69} & \ncell{562} \\
        \lightrow 
        \textit{Rbot3} & \ncell{80327} & \ncell{31388} & \ncell{98.664} & \ncell{33} & \ncell{816} & \ncell{-3} & \ncell{795} \\
        \darkrow 
        \textit{Rbot4} & \ncell{4573} & \ncell{3073} & \ncell{99.854} & \ncell{0} & \ncell{9} & \ncell{0} & \ncell{9} \\
        \lightrow 
        \textit{Virut1} & \ncell{3807} & \ncell{306} & \ncell{74.568} & \ncell{4} & \ncell{202} & \ncell{-18} & \ncell{140} \\
        \yellowrow 
        \textit{Virut2} & \ncell{70762} & \ncell{8714} & \ncell{87.040} & \ncell{148} & \ncell{2403} & \ncell{-1703} & \ncell{1003} \\
        \greenrow
        Summary & \ncell{671656} & \ncell{101261} & \ncell{90.803} & \ncell{597} & \ncell{19795} & \ncell{-2977} & \ncell{14596}
    \end{tabular}
    \end{tabular}
    }
    \label{tab:Ensemble}
\end{table*}

Let us discuss Table~\ref{tab:Ensemble}, starting from the CICIDS2017 results. Here, we highlight (in yellow) \emph{Botnet1} and \emph{DoS slowhttptest} scenarios because they show the greatest improvement of the FNN+LSTM ensemble over the stand-alone LSTM. In these scenarios, the ensemble has lower \textit{fn} than the LSTM, and the same \textit{fp} (only 1 more for the \textit{Botnet1}).
This is possible because the FNN used in the ensemble correctly identifies the malicious NetFlows that are misclassified by the LSTM.
We observe an interesting phenomenon in four scenarios, i.e., \emph{DoS slowhttptest}, \emph{Botnet1-2} and \emph{Port Scan}. The ensemble has not only lower \textit{fn} than the stand-alone LSTM, but also lower than the stand-alone FNN (see Table~\ref{tab:LSTM_FNNResults}).
This means the ensemble also benefits from the LSTM because it correctly identifies malicious NetFlows misclassified by the FNN.

We now focus on the CTU13. Here, the FNN+LSTM ensemble does better than the stand-alone LSTM in terms of number of false-negatives. However, the latter has a higher F1-score, i.e.,  $95.703\%$ vs $90.803\%$.
This is due to the more conservative nature of the ensemble in detecting malicious events, which may reduce the \textit{fn}, but may increase the \textit{fp}.
A more in-depth analysis of each scenario outlines that the ensemble reduces the \textit{fn} in almost all cases. Notably, in \emph{Neris1} and \emph{Virut2} scenarios (in yellow) the FNN+LSTM ensemble has significantly lower \textit{fn} than the stand-alone LSTM, but it also has over $2$k more \textit{fp}. The rationale is that the FNN is better in detecting malicious NetFlows, which benefits the FNN+LSTM ensemble. However, we also observe that the ensemble also benefits from the LSTM. This is clear if we compare the overall number of false-negatives of the FNN+LSTM ensemble with those of the stand-alone FNN (see Table~\ref{tab:LSTM_FNNResults}): the former has only 597, while the latter has almost four times that amount with 1947.

\vspace{0.5em}
\takeaway{the LSTM and FNN appear to have learned different classification patterns, confirming that the application of our method to sequential ML models \textit{does} influence their detection.}

\section{Conclusions}
\label{sec:conclusions}

In this work, we investigate the use of Long Short-Term Memory (LSTM) neural networks to learn temporal patterns among NetFlows as a result of cyber-attacks.
We review and highlight the limitations of related work, which has not truly studied the effectiveness of temporal patterns for NetFlow-based NIDS. Hence, we propose an original method that can be used to extract temporal patterns from labelled NetFlow data. We then apply this method to train a `sequential' LSTM classifier, and compare its performance against a `static' Feedforward Neural Network (FNN) to verify if temporal patterns are truly significant for ML-NIDS.

Our evaluation spans over two recent ML-NIDS datasets, CICIDS2017 and CTU13. The results highlight that LSTM achieves comparable ($\sim \! 99\%$ F1-score) or better ($95\%$ vs $91\%$ F1-score) performance in detecting malicious NetFlows than the FNN. However, against some specific attacks, the LSTM proves to be significantly better, yielding lower false negatives and lower false positives.

We verify if the proposed method influences the detection by creating an ensemble of the FNN+LSTM models, and compare it against the stand-alone LSTM. This additional set of experiments further confirms that the LSTM and FNN models learn different classification patterns, which translate to different performance that can be exploited to mitigate some types of attacks. Such comparison confirms that, if properly extracted (e.g., with the proposed method and sequential ML-models), temporal patterns can truly be effective at enhancing the performance of ML-NIDS.

Our paper paves the way to more secure and effective NIDS, which rely not only on `static' ML methods, but also by sequential approaches that leverage the underlying relationships among distinct network traffic samples.



\appendix
\section{Appendix A}
\label{app:hyperparameters}

When preprocessing the datasets, we one-hot encode the categorical features: source/destination ports, the direction and the protocol. The numerical features are normalized in the range $[0, 1]$. We grid-search the most optimal parameters of each model, which we report in Table~\ref{tab:hyperparams}).

\begin{table}[!htbp]
\centering
\caption{Hyperparameters (CICIDS2017: left; CTU13: right).}
\arrayrulecolor{black}
    \resizebox{0.85\columnwidth}{!}{
    \begin{tabular}{c|c|c||c|c}
    \toprule
    \textbf{parameter}
    &\textbf{\textit{LSTM}} 
    &\textbf{\textit{FNN}}
    &\textbf{\textit{LSTM}}
    &\textbf{\textit{FNN}}\\
    \midrule
    \midrule
    learning rate & 0.001 & 0.001 & 0.001 & 0.001 \\
    \# epochs & 30 & 20 & 100 & 100 \\
    dropout & 0.2 & 0.3 & 0.2 & 0.2 \\
    batch size & 1 & 512 & 1 & 1024 \\
    optimizer & Adam & Adam & Adam & Adam \\
    truncated BPTT window & N/A & N/A & 512 & N/A \\
    LSTM1 size & 256 & N/A & 256 & N/A \\
    LSTM2 size & 256 & N/A & 256 & N/A \\
    FC1 size & N/A & 256 & N/A & 512 \\
    FC2 size & N/A & 256 & N/A & 512 \\
    FC size & 2 & 2 & 2 & 2 \\
    \bottomrule
    \end{tabular}
    }
\label{tab:hyperparams}
\end{table}

\end{document}